\documentclass[submission,copyright,creativecommons]{eptcs}
\providecommand{\event}{MeCBIC 2012} 
\usepackage{breakurl}             
\usepackage{graphs}
\usepackage{graphicx}
\usepackage{latexsym}
\usepackage{amsfonts}
\usepackage{amssymb}
\usepackage{amsmath}
\usepackage{mathptmx}

\newtheorem{theorem}{Theorem}

\newtheorem{problem}{Problem}

\newcommand{\xvec}     {\mathbf{x}}
\newcommand{\yvec}     {\mathbf{y}}
\newcommand{\zvec}     {\mathbf{z}}
\newcommand{\pvec}     {\mathbf{p}}
\newcommand{\RS}       {\mathit{regSteps}}
\newcommand{\BMS}      {{\mathit{BMS}}}
\newcommand{\PTL}      {{\mathit{PTL}}}
\newcommand{\AS}       {\mathit{tsSteps}}
\newcommand{\CRG}      {\mathit{CRG}}
\newcommand{\TS}       {\mathit{TS}}

\newcommand{\Mode}        {\mathfrak{m}}
\newcommand{\lhs}         {\mathit{lhs}}
\newcommand{\rhs}         {\mathit{rhs}}
\newcommand{\ie}          {i.e., }
\newcommand{\eg}          {e.g., }

\newcommand{\es}          {\varnothing}
\newcommand{\pre}[1]      {{}^\bullet{#1}}
\newcommand{\post}[1]     {{{#1}}^\bullet}
\newcommand{\STEP}[1]     {[#1\rangle}
\newcommand{\nsymbol}     {{\mathbb{N}}}
\newcommand{\goesONE}[1]  {\stackrel{#1}{\longrightarrow}}

\newcommand{\StandardSize} {\setlength{\unitlength}{0.65cm}}

\newcommand{\StandardNet}    {\graphnodecolour{1}
                              \graphnodesize{1.1}
                              \grapharrowlength{0.2}
                              \graphlinewidth{0.02}
                              \grapharrowwidth{0.7}
                              \autodistance{1.15}
                              }

\newcommand{\SOUTH} [2]    {\autonodetext{#1}[s]{$ #2$}}
\newcommand{\NORTH} [2]    {\autonodetext{#1}[n]{$ #2$}}
\newcommand{\WEST} [2]     {\autonodetext{#1}[w]{$ #2$}}
\newcommand{\EAST} [2]     {\autonodetext{#1}[e]{$ #2$}}

\newcommand{\place}[4]     {\roundnode{#1}(#2,#3)%
                             \freetext(#2,#3){$#4$}[\opaquetextfalse]}
\newcommand{\placE}[5]     {\place{#1}{#2}{#3}{#4}\EAST{#1}{ #5}}
\newcommand{\placW}[5]     {\place{#1}{#2}{#3}{#4}\WEST{#1}{ #5}}
\newcommand{\placN}[5]     {\place{#1}{#2}{#3}{#4}\NORTH{#1}{ #5}}
\newcommand{\placS}[5]     {\place{#1}{#2}{#3}{#4}\SOUTH{#1}{ #5}}

\newcommand{\Aplace}[4]     {\roundnode{#1}(#2,#3)[\graphnodecolour{.98}]%
                             \freetext(#2,#3){$#4$}[\opaquetextfalse]}

\newcommand{\AplacN}[5]     {\Aplace{#1}{#2}{#3}{#4}\NORTH{#1}{ #5}}

\newcommand{\Bplace}[4]     {\roundnode{#1}(#2,#3)[\graphnodecolour{.65}]%
                             \freetext(#2,#3){$#4$}[\opaquetextfalse]}

\newcommand{\BplacS}[5]     {\Bplace{#1}{#2}{#3}{#4}\SOUTH{#1}{ #5}}

\newcommand{\Cplace}[4]     {\roundnode{#1}(#2,#3)[\graphnodecolour{.9}]%
                             \freetext(#2,#3){$#4$}[\opaquetextfalse]}
\newcommand{\CplacE}[5]     {\Cplace{#1}{#2}{#3}{#4}\EAST{#1}{ #5}}
\newcommand{\CplacW}[5]     {\Cplace{#1}{#2}{#3}{#4}\WEST{#1}{ #5}}

\newcommand{\trans}[4]     {\squarenode{#1}(#2,#3)%
                              \freetext(#2,#3){$#4$}[\opaquetextfalse]}
\newcommand{\transA}[4]    {\squarenode{#1}(#2,#3)[\graphnodecolour{.98}]%
                              \freetext(#2,#3){$#4$}[\opaquetextfalse]}
\newcommand{\transB}[4]    {\squarenode{#1}(#2,#3)[\graphnodecolour{.65}]%
                              \freetext(#2,#3){$#4$}[\opaquetextfalse]}
\newcommand{\transC}[4]    {\squarenode{#1}(#2,#3)[\graphnodecolour{.9}]%
                              \freetext(#2,#3){$#4$}[\opaquetextfalse]}

\newcommand{\TwoDOTS}{%
\StandardSize
\begin{graph}(0,0)
    \StandardNet
    \roundnode{A}(-0.1, 0.15)[\graphnodesize{0.2}\graphnodecolour{0}]
    \roundnode{B}(0.1,-0.15)[\graphnodesize{0.2}\graphnodecolour{0}]
\end{graph}}

\newcommand{\DOT}{%
\StandardSize
\begin{graph}(0,0)
    \StandardNet
    \roundnode{P}(0,0)[\graphnodesize{0.2}\graphnodecolour{0}]
\end{graph}}

\newcommand{\BUBBLE}[6]
{\bubble{#1}{(#2,#4)(#2,#5)(#3,#5)(#3,#4)}[\graphfillcolour{#6}\graphlinecolour{#6}\graphlinewidth{0}]}

\title{Membrane Systems and Petri Net Synthesis\\(Invited Paper)}
\author{
\makebox[5cm][c]{Jetty Kleijn}
\institute{LIACS\\Leiden University\\Leiden, The Netherlands}
\email{kleijn@liacs.nl}
\and
\makebox[5cm][c]{Maciej Koutny}
\institute{School of Computing Science\\Newcastle University\\Newcastle upon Tyne, UK}
\email{maciej.koutny@ncl.ac.uk}
\and
\makebox[5cm][c]{Marta Pietkiewicz-Koutny}
\institute{School of Computing Science\\Newcastle University\\Newcastle upon Tyne, UK}
\email{marta.koutny@ncl.ac.uk}
\and
\makebox[5cm][c]{Grzegorz Rozenberg}
\institute{LIACS\\Leiden University\\Leiden, The Netherlands\\and\\
Department of Computer Science\\University of Colorado at Boulder
\\Boulder, Colorado, USA}
\email{rozenber@liacs.nl}
}
\def\titlerunning{Membrane Systems and Petri Net Synthesis}
\def\authorrunning{J.Kleijn, M.Koutny, M.Pietkiewicz-Koutny \& G.Rozenberg}

\begin{document}
\maketitle

\begin{abstract}
Automated synthesis from behavioural specifications
is an attractive and powerful way of constructing concurrent systems.
Here we focus on the problem of synthesising
a membrane system from a behavioural specification
given in the form of a transition system
which specifies the desired state space of the system to be constructed.
We demonstrate how a Petri net solution to this problem,
based on the notion of region of a transition system,
yields a method of automated synthesis of membrane
systems from state spaces.
\end{abstract}

\section{Introduction}

Membrane systems (\cite{Pa-2000,Pa-2002,PR-2002,PRS-2009})
are a computational model inspired by
the functioning of living cells and their architecture
and in particular,
the way chemical reactions take place in cells
divided by membranes into compartments.
The reactions are abstracted to rules that specify
which and how many molecules can be produced from given
molecules of a certain kind and quantity. As a result,
membrane systems
are essentially multiset rewriting systems.
The dynamic aspects of the membrane system model
including potential behaviour (computations),
derive from such evolution rules.

Petri nets (see, \eg \cite{desel-juhas,petri-62,RR98})
are a well-established general model for distributed computation
with an extensive range of tools and methods
for construction, analysis, and verification of concurrent systems.
Their diverse applications areas
include computational and operational foundations
for problems and issues arising in biology; see
for example, \cite{KRS2011}, for a recent comprehensive
overview of  applications of Petri
nets in systems biology.

There are intrinsic similarities between Petri nets and membrane systems.
In particular, there exists a canonical way of
translating membrane systems into Petri nets.
This translation is faithful in the sense that it relates computation steps at
the lowest level and induces in a natural way (sometimes new) extensions and
interpretations of Petri net structure and behaviour (e.g., inhibitor arcs,
localities, and maximal concurrency).
More details on the relationship between Petri nets and membrane
systems can be found in, \eg~\cite{our-handbook,KKR11}.

The strong semantical link between the two models invites to extend where
necessary and possible existing Petri net techniques and bring them to the domain of
membrane systems. An example is the process semantics of Petri nets
that can help to understand the dynamics and causality
in the biological evolutions represented by membrane systems~\cite{KK-TCS,KKR06}.
In this paper, we focus on the synthesis problem,
that is, the problem of automated construction of a system
from a specification
of its (observed or desired) behaviour.

Automated synthesis from behavioural specifications
is an attractive and powerful way of
constructing correct concurrent systems~\cite{BD97,Ber93,DKPKY08,DR-96,ER90,M92,MPK99}.
Here we will re-visit the problem of synthesising
a Petri net from a behavioural specification
given in the form of a transition system. The latter
specifies the desired state space of the Petri net to be constructed.
We will recall a solution to this problem
based on the notion of region of a transition system.
We will then demonstrate how
this solution
leads to a method of automated synthesis of basic membrane
systems from state spaces.
We also discuss how the proposed method could be extended to
cope with more complicated kinds of membrane systems.

\section{Preliminaries}

\paragraph{Multisets.}

A multiset over a
finite set $X$ is
a function $\theta:X\to\nsymbol=\{0,1,2,\ldots\}$.
$\theta$ may be represented by listing its elements with repetitions,
\eg $\theta=\{y,y,z\}$ is such that $\theta(y)=2$, $\theta(z)=1$,
and $\theta(x)=0$ otherwise.
$\theta$ is said to be empty (and denoted by $\es$) if there are no $x$ such that
$x\in \theta$ by which we mean that $x\in X$ and $\theta(x)\geq 1$.

For two multisets $\theta$ and $\theta'$ over $X$, the
sum $\theta+\theta'$ is the multiset given by
$(\theta+\theta')(x)=\theta(x)+\theta'(x)$ for all $x\in X$,
and for $k\in\nsymbol$ the multiset $k\cdot\theta$ is given by
$(k\cdot\theta)(x)= k\cdot\theta(x)$ for all $x\in X$.
The difference $\theta-\theta'$ is given by
$(\theta-\theta')(x)=\max\{\theta(x)-\theta'(x),0\}$ for all $x\in X$.
We denote $\theta\leq\theta'$ whenever $\theta(x)\leq\theta'(x)$ for all $x\in X$, and
$\theta<\theta'$ whenever $\theta\leq\theta'$ and $\theta\neq \theta'$.
The restriction $\theta|_Z$ of $\theta$ to a subset $Z\subseteq X$ is given
by $\theta|_Z(x)=\theta(x)$ for $x\in Z$, and $\theta|_Z(x)=0$ otherwise.
The size $|\theta|$ of $\theta$ is given by $\sum_{x\in X} \theta(x)$.
If $f:X\to Y$ is a function then $f(\theta)$ is the multiset over $Y$ such
that $f(\theta)(y)=\sum_{x\in f^{-1}(y)}\theta (x)$, for every $y\in Y$.

\paragraph{Step transition systems.}

A \emph{step transition system} over a
finite
set (of actions) $A$ is a triple
$\TS=(Q,\mathcal{A}, q_0)$, where:
$Q$ is a set of nodes called \emph{states};
$\mathcal{A}$ is the set of \emph{arcs}, each arc being a triple
$(q,\alpha,q')$ such that $q,q'\in Q$ are states and
$\alpha$ is a
multiset over $A$; and $q_0\in Q$ is the \emph{initial} state.
We may write $q\xrightarrow{\alpha}q'$
whenever $(q,\alpha,q')$ is an arc, and denote by
\[
    \AS_q  =\{\alpha \mid \alpha\neq\es~\wedge~\exists q': ~q\xrightarrow{\alpha}q'\}
\]
the set of nonempty steps enabled at a state $q$ in $\TS$.
We additionally assume that:
\begin{itemize}
\item
    if $q\xrightarrow{\alpha}q'$ and $q\xrightarrow{\alpha}q'' $
    then $q'=q''$ (\ie $\TS$ is deterministic);

\item
    for every state $q\in Q$, there is a path from $q_0$ leading to $q$;

\item
    for every action $a\in A$, there is an arc $q\xrightarrow{\alpha}q'$ in $\TS$
    such that $a\in\alpha$;
and
\item
    for every state $q\in Q$, we have $q\xrightarrow{\es}q'$ iff $q=q'$.
\end{itemize}

Let $\TS=(Q,\mathcal{A}, q_0)$ be a step transition system over a set of actions $A$, and
$\TS'=(Q',\mathcal{A}', q'_0)$ be a step transition system over a set of actions  $A'$.
$\TS$ and $\TS'$ are \emph{isomorphic} if there are two bijections,
$\phi:A\to A'$ and $\nu: Q\to Q'$, such that $\nu(q_0)=q'_0$ and, for all states $q,q'\in Q$ and
multisets $\alpha$ over $A$:
\[
    (q,\alpha,q')\in \mathcal{A}
    ~~~
    \Longleftrightarrow
    ~~~
    (\nu(q),\phi(\alpha),\nu(q'))\in \mathcal{A}' \;.
\]
We denote this by $\TS\sim_{\phi,\nu}\TS'$ or $\TS\sim\TS'$.

\paragraph{Petri nets.}
\label{section-PNs}

\begin{figure}[!ht]
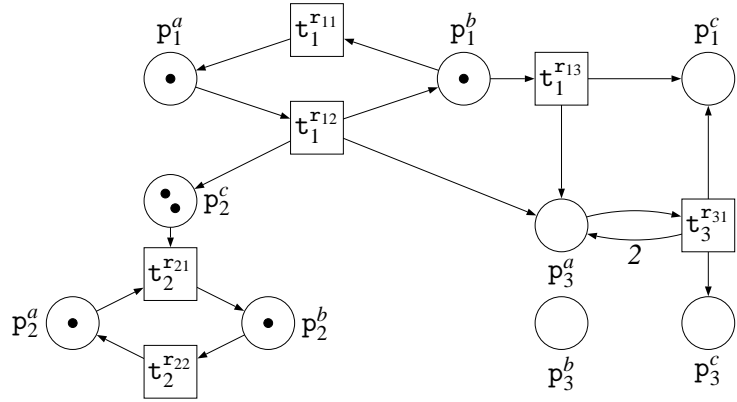

\begin{center}
\StandardSize

\begin{graph}(18,9)
\StandardNet

\opaquetextfalse

    \placN{A01}{ 5}{7}{\DOT}{\mathtt{p}^a_1}
    \placN{B01}{11}{7}{\DOT}{\mathtt{p}^b_1}
    \placN{C01}{16}{7}{}{\mathtt{p}^c_1}

    \trans{R14}{13}{7}{\mathtt{t}^{\mathtt{r}_{13}}_1}
    \trans{R11}{ 8}{8}{\mathtt{t}^{\mathtt{r}_{11}}_1}
    \trans{R12}{ 8}{6}{\mathtt{t}^{\mathtt{r}_{12}}_1}

    \diredge{A01}{R12}
    \diredge{R12}{B01}
    \diredge{B01}{R11}
    \diredge{R11}{A01}
    \diredge{B01}{R14}
    \diredge{R14}{C01}

    \diredge{R12}{C02}
    \diredge{R12}{A03}

    \placW{A02}{ 3}{2}{\DOT}{\mathtt{p}^a_2}
    \placE{B02}{ 7}{2}{\DOT}{\mathtt{p}^b_2}
    \placE{C02}{ 5}{4.5}{\TwoDOTS}{\mathtt{p}^c_2}

    \trans{R21}{ 5}{3}{\mathtt{t}^{\mathtt{r}_{21}}_2}
    \trans{R22}{ 5}{1}{\mathtt{t}^{\mathtt{r}_{22}}_2}

    \diredge{R22}{A02}
    \diredge{B02}{R22}
    \diredge{R21}{B02}
    \diredge{A02}{R21}
    \diredge{C02}{R21}

    \placS{A03}{13}{4}{}{\mathtt{p}^a_3}
    \placS{B03}{13}{2}{}{\mathtt{p}^b_3}
    \placS{C03}{16}{2}{}{\mathtt{p}^c_3}

    \trans{R31}{16}{4}{\mathtt{t}^{\mathtt{r}_{31}}_3}

    \dirbow{A03}{R31}{.1}
    \dirbow{R31}{A03}{.1}\bowtext{R31}{A03}{.2}{$\mathit{2}$}
    \diredge{R31}{C01}

    \diredge{R14}{A03}
    \diredge{R31}{C03}
\end{graph}

\end{center}
\caption{
    \label{FIG-333}
    A \textsc{pt}-net.
    }
\end{figure}

A \emph{Place/Transition net}
(or \textsc{pt}-net) is specified as a tuple
$\mathit{PT}= (P,T,W,M_0)$,
where:
$P$ and $T$ are finite
disjoint sets of respectively \emph{places} and
\emph{transitions};
$W:(T\times P)\cup(P\times T)\rightarrow\nsymbol$
is the \emph{arc weight function};
and
$M_0:P\rightarrow\nsymbol$ is the \emph{initial marking}
(in general, any multiset of places is a marking).
We assume that, for each transition $t$,
there is at least one place $p$ such that $W(p,t)>0$.
In diagrams, such as that in Figure~\ref{FIG-333},
places are drawn as circles,
and transitions as boxes.
If $W(x,y) \geq 1$,
then $(x,y)$ is an \emph{arc} leading from $x$ to $y$.
An arc is annotated
with its weight if the latter is greater than one.
A marking $M$ is
represented by drawing in each place $p$ exactly $M(p)$ tokens
(small black dots).

A \emph{step} $U$ of $\mathit{PT}$ is a multiset of transitions.
Its \emph{pre-multiset} and \emph{post-multiset} of places,
$\pre{U}$ and $\post{U}$, are respectively given by
\[
    \pre{U}(p)=\sum_{t\in U}U(t)\cdot W(p,t)
    ~~~\mbox{and}~~~
    \post{U}(p)=\sum_{t\in U}U(t)\cdot  W(t,p)\;,
\]
for each place $p$.
For the \textsc{pt}-net in Figure~\ref{FIG-333} we have:
\[
    \pre{\{\mathtt{t}^{\mathtt{r}_{11}}_1,\mathtt{t}^{\mathtt{r}_{11}}_1,
    \mathtt{t}^{\mathtt{r}_{31}}_3\}}
    =
    \{\mathtt{p}^b_1,\mathtt{p}^b_1,\mathtt{p}^a_3\}
    ~\mbox{and}~
    \post{\{\mathtt{t}^{\mathtt{r}_{11}}_1,\mathtt{t}^{\mathtt{r}_{11}}_1,\mathtt{t}^{\mathtt{r}_{31}}_3\}}
    =
    \{\mathtt{p}^a_1,\mathtt{p}^a_1,\mathtt{p}^c_1,\mathtt{p}^a_3,\mathtt{p}^a_3,\mathtt{p}^c_3\}.
\]

We distinguish two basic modes of execution of \textsc{pt}-nets.
To start with, a step of transitions $U$ is
\emph{free-enabled}
at a marking $M$ if $\pre{U}\leq M$.
We denote this by $M\STEP{U}_\mathit{free}$, and then say that
a free-enabled $U$ is
\emph{max-enabled} at $M$ if $U$ cannot be extended by a transition
to yield a step which is free-enabled at $M$,
\ie there is no $t\in T$ such that $M\STEP{U+\{t\}}_\mathit{free}$.
We denote this by $M\STEP{U}_\mathit{max}$.
In other words,
$U$ is free-enabled at $M$ if
in each place there are sufficiently many tokens for the specified multiple occurrence
of each of its transitions.
Maximal concurrency (max-enabledness) means that extending $U$  would
demand more tokens than $M$ supplies.
For the \textsc{pt}-net in Figure~\ref{FIG-333} we have that, at the
given marking $M_0$, the step
$\{\mathtt{t}^{\mathtt{r}_{12}}_1,\mathtt{t}^{\mathtt{r}_{21}}_2\}$ is
free-enabled but not max-enabled, and
$\{\mathtt{t}^{\mathtt{r}_{11}}_1,\mathtt{t}^{\mathtt{r}_{12}}_1,
\mathtt{t}^{\mathtt{r}_{21}}_2,\mathtt{t}^{\mathtt{r}_{22}}_2\}$ is max-enabled.

For each mode of execution $\Mode\in\{\mathit{free}, \mathit{max}\}$,
a step $U$ which is $\Mode$-enabled at a marking $M$
can be \emph{$\Mode$-executed} leading
to the marking $M'$ given by
$M'= M-\pre{U}+\post{U}$.
We denote this by $M\STEP{U}_\Mode M'$.
For the \textsc{pt}-net in Figure~\ref{FIG-333} we have
\[
    M_0\STEP{\{\mathtt{t}^{\mathtt{r}_{12}}_1,
    \mathtt{t}^{\mathtt{r}_{21}}_2\}}_\mathit{free}
    \{\mathtt{p}^b_1,\mathtt{p}^b_1,\mathtt{p}^b_2,
    \mathtt{p}^b_2,\mathtt{p}^c_2,\mathtt{p}^c_2,\mathtt{p}^a_3\}\;.
\]

\paragraph{Petri nets with localities.}
\label{section-PNLs}

\textsc{pt}-nets are a general model of concurrent computation.
To capture the compartmentisation of membrane systems,~\cite{KKR06}
adds explicit localities to transitions.
Though not necessary from a modelling point of view,
we associate in this paper --- only for notational
convenience --- also each place with a locality.

\begin{figure}[!ht]
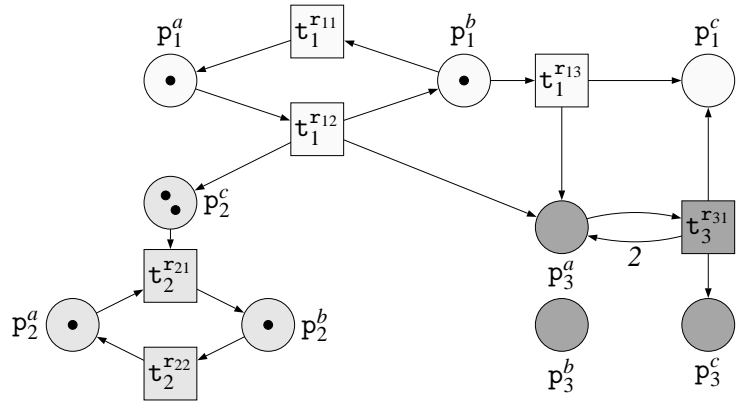

\begin{center}
\StandardSize

\begin{graph}(18,9)(1,0)
\StandardNet

\opaquetextfalse

    \AplacN{A01}{ 5}{7}{\DOT}{\mathtt{p}^a_1}
    \AplacN{B01}{11}{7}{\DOT}{\mathtt{p}^b_1}
    \AplacN{C01}{16}{7}{}{\mathtt{p}^c_1}

    \transA{R14}{13}{7}{\mathtt{t}^{\mathtt{r}_{13}}_1}
    \transA{R11}{ 8}{8}{\mathtt{t}^{\mathtt{r}_{11}}_1}
    \transA{R12}{ 8}{6}{\mathtt{t}^{\mathtt{r}_{12}}_1}

    \diredge{A01}{R12}
    \diredge{R12}{B01}
    \diredge{B01}{R11}
    \diredge{R11}{A01}
    \diredge{B01}{R14}
    \diredge{R14}{C01}

    \diredge{R12}{C02}
    \diredge{R12}{A03}

    \CplacW{A02}{ 3}{2}{\DOT}{\mathtt{p}^a_2}
    \CplacE{B02}{ 7}{2}{\DOT}{\mathtt{p}^b_2}
    \CplacE{C02}{ 5}{4.5}{\TwoDOTS}{\mathtt{p}^c_2}

    \transC{R21}{ 5}{3}{\mathtt{t}^{\mathtt{r}_{21}}_2}
    \transC{R22}{ 5}{1}{\mathtt{t}^{\mathtt{r}_{22}}_2}

    \diredge{R22}{A02}
    \diredge{B02}{R22}
    \diredge{R21}{B02}
    \diredge{A02}{R21}
    \diredge{C02}{R21}

    \BplacS{A03}{13}{4}{}{\mathtt{p}^a_3}
    \BplacS{B03}{13}{2}{}{\mathtt{p}^b_3}
    \BplacS{C03}{16}{2}{}{\mathtt{p}^c_3}

    \transB{R31}{16}{4}{\mathtt{t}^{\mathtt{r}_{31}}_3}

    \dirbow{A03}{R31}{.1}
    \dirbow{R31}{A03}{.1}\bowtext{R31}{A03}{.2}{$\mathit{2}$}
    \diredge{R31}{C01}

        \diredge{R14}{A03}
    \diredge{R31}{C03}
\end{graph}

\end{center}
\caption{
    A \textsc{ptl}-net corresponding to a basic membrane system,
    where $\ell(x^z_i)=i$, for each node of the form $x^z_i$.
    Note that, \eg transitions
    $\mathtt{t}^{\mathtt{r}_{11}}_1$,
    $\mathtt{t}^{\mathtt{r}_{12}}_1$ and
    $\mathtt{t}^{\mathtt{r}_{13}}_1$ are co-located.
    }
    \label{FIG-3aa}
\end{figure}

A \emph{\textsc{pt}-net with localities}
(or \textsc{ptl}-net) is a tuple
$\PTL = (P,T,W,\ell,M_0)$ such that
$(P,T,W,M_0)$ is a \textsc{pt}-net,
and $\ell$ is a
\emph{location} mapping
for the transitions and places.
Whenever $\ell(x)=\ell(z)$, we call $x$ and $z$ \emph{co-located}.
In diagrams,
nodes representing co-located transitions and/or places will be shaded in the same way,
as shown in
Figure~\ref{FIG-3aa}.

Co-locating transitions leads to one more way of enabling for steps of
transitions. We say that a \emph{step} $U$ of $\PTL $ is
\emph{lmax-enabled} at a marking $M$ if $M\STEP{U}_\mathit{free}$ and $U$
cannot be extended by a transition co-located with a transition in $U$
to yield a step which is free-enabled at~$M$;
\ie there is no $t\in T$ such that
$\ell(t)\in\ell(U)$ and $M\STEP{U+\{t\}}_\mathit{free}$. We denote this
by $M\STEP{U}_\mathit{lmax}$, and then denote the lmax-execution of $U$
by $M\STEP{U}_\mathit{lmax}M'$, where $M'= M-\pre{U}+\post{U}$.
Note that
\emph{locally maximal (lmax) concurrency}  is similar to maximal
concurrency, but now only
active localities\footnote
{
By active localities of a step $U$ we mean the localities of
transitions present in $U$.
}
cannot execute further transitions.
For the \textsc{ptl}-net in Figure~\ref{FIG-3aa} we have that
$\{\mathtt{t}^{\mathtt{r}_{11}}_1,\mathtt{t}^{\mathtt{r}_{12}}_1\}$
is lmax-enabled at the given marking,
but $\{\mathtt{t}^{\mathtt{r}_{11}}_1\}$ is not.

Let $\Mode\in\{\mathit{free},\mathit{max},\mathit{lmax}\}$ be a mode
of execution of a \textsc{ptl}-net $\PTL$.
Then an \emph{$\Mode$-step sequence} is a finite sequence of $\Mode$-executions
starting from the initial marking,
and an \emph{$\Mode$-reachable} marking is any marking resulting from the execution of
such a sequence.
Moreover,
the \emph{$\Mode$-concurrent reachability graph of $\PTL$} is
the step transition system:
\[
    \CRG_\Mode(\PTL)
    =
    \Big(~\STEP{M_0}_\Mode~,~
    \big\{(M,U,M')
    \mid M\in \STEP{M_0}_\Mode\;\wedge\; M\STEP{U}_\Mode M'
    \big\}~,~M_0~\Big) \;,
\]
where $\STEP{M_0}_\Mode$ is the set of all $\Mode$-reachable markings
which are the nodes of the graph; $M_0$ is the initial node;
and the arcs between the nodes are labelled by $\Mode$-executed steps
of transitions.
Concurrent reachability graphs provide complete representations
of the dynamic behaviour of \textsc{ptl}-nets
evolving according to the chosen mode of execution.

\paragraph{Membrane structures.}
\label{sect-section-2}

\begin{figure}[!t]
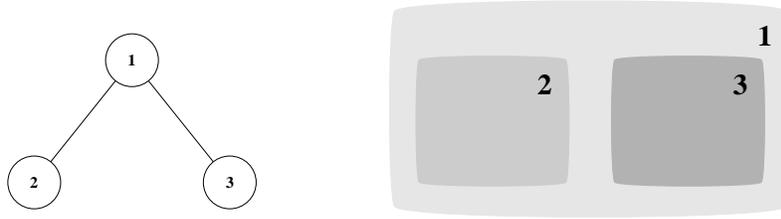

\begin{center}
\StandardSize

\begin{graph}(4,4.5)
\StandardNet
\opaquetextfalse

\placE{A1}{2}{3.0}{\mbox{{\tiny\bf 1}}}{}
\placE{A2}{0}{0.5}{\mbox{{\tiny\bf 2}}}{}
\placN{A3}{4}{0.5}{\mbox{{\tiny\bf 3}}}{}

\edge{A1}{A2}\edge{A1}{A3}
\end{graph}
\hspace*{2cm}
\begin{graph}(8,4)
\StandardNet

\BUBBLE{.05}{0}{8}{0}{4}{.9}
\textnode{X1}(7.5,3.5){{ $\mathbf{1}$}}[\graphlinewidth{0}\graphnodecolour{.9}\graphlinecolour{.9}]

\BUBBLE{.05}{.5}{3.5}{.5}{3}{.8}
\textnode{X2}(3,2.5){{ $\mathbf{2}$}}[\graphlinewidth{0}\graphnodecolour{.8}\graphlinecolour{.8}]

\BUBBLE{.05}{4.5}{7.5}{.5}{3}{.7}
\textnode{X3}(7,2.5){{ $\mathbf{3}$}}[\graphlinewidth{0}\graphnodecolour{.7}\graphlinecolour{.7}]

\end{graph}
\end{center}
\caption{
    A membrane structure ($m=3$) and its compartments with  1
    being the root node, $(1,2)\in\mu$ and $1=\mathit{parent}(3)$.
    }
    \label{FIG-1}
\end{figure}

A \emph{membrane structure} $\mu$ (of degree $m\geq 1$) is given by a
rooted tree with $m$ nodes identified with the integers $1,\ldots,m$. We
will write $(i,j)\in \mu$ or $i=\mathit{parent}(j)$ to indicate
that there is
an edge from $i$ (parent) to $j$ (child) in the tree of~$\mu$, and
$i\in\mu$ means that $i$ is a node of $\mu$. The nodes of a membrane
structure represent nested membranes which in turn determine compartments
(compartment $i$ is enclosed by membrane $i$ and lies in-between $i$ and
its children, if any), as shown in Figure~\ref{FIG-1}.

We will say that a \textsc{ptl}-net
$\PTL=(P,T,W,\ell,M_0)$ is \emph{spanned} over the membrane structure $\mu$
if $\ell:P\cup T\to\mu$ and the following hold, for all $p\in P$ and  $t\in T$:
\begin{itemize}
\item
    if $W(p,t)>0$ then $\ell(p)=\ell(t)$; and

\item
    if $W(t,p)>0$ then
    $\ell(p)=\ell(t)$ or
    $(\ell(p),\ell(t))\in\mu$ or
    $(\ell(t),\ell(p))\in\mu$.
\end{itemize}
The \textsc{ptl}-net of Figure~\ref{FIG-3aa} is spanned over the membrane structure depicted
in Figure~\ref{FIG-1}.

\begin{figure}[!ht]
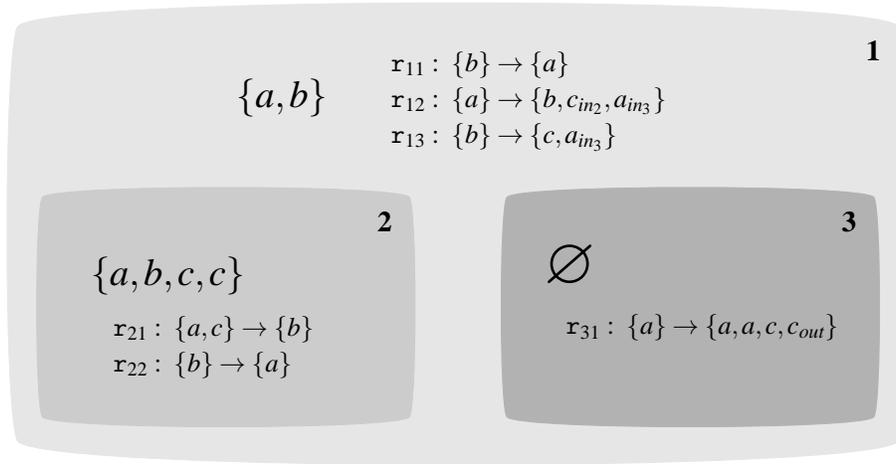

\begin{center}
\StandardSize
\begin{graph}(18,10)(1,0)
\StandardNet

\BUBBLE{.05}{1}{19}{0}{8.5}{.9}
\textnode{X1}(18.5,8){{$\mathbf{1}$}}[\graphlinewidth{0}\graphnodecolour{.9}\graphlinecolour{.9}]

\BUBBLE{.05}{1.5}{9}{.5}{5}{.8}
\textnode{X2}(8.5,4.5){{$\mathbf{2}$}}[\graphlinewidth{0}\graphnodecolour{.8}\graphlinecolour{.8}]

\BUBBLE{.05}{11}{18.5}{.5}{5}{.7}
\textnode{X3}(18,4.5){{$\mathbf{3}$}}[\graphlinewidth{0}\graphnodecolour{.7}\graphlinecolour{.7}]

\opaquetextfalse

\textnode{A}(10,7)
    {$ \mbox{\Large$\{a,b\}$}~~~~~~~
    \begin{array}{l}
    \mbox{\small $\mathtt{r}_{11}:~\{b\}\rightarrow \{a\}$}
    \\
    \mbox{\small $\mathtt{r}_{12}:~\{a\}\rightarrow \{b,c_{\mathit{in}_2},a_{\mathit{in}_3}\}$}
    \\
    \mbox{\small $\mathtt{r}_{13}:~\{b\}\rightarrow  \{c,a_{\mathit{in}_3}\}$}
    \end{array}
    $}[\graphlinewidth{0}\graphnodecolour{.9}\graphlinecolour{.9}]

\textnode{B}(5,2.5)
    {$
    \begin{array}{l}
    \mbox{\hspace{-.3cm}\Large$\{a,b,c,c\}$}
    \\[2mm]
    \mbox{\small $\mathtt{r}_{21}:~\{a,c\}\rightarrow \{b\}$}
    \\
    \mbox{\small $\mathtt{r}_{22}:~\{b\}\rightarrow \{a\}$}
    \end{array}
    $}[\graphlinewidth{0}\graphnodecolour{.8}\graphlinecolour{.8}]

\textnode{C}(15,3)
    {$
    \begin{array}{l}
    \mbox{\hspace{-.3cm}\Huge$\es$}
    \\[2mm]
    \mbox{\small $\mathtt{r}_{31}:~\{a\}\rightarrow \{a,a,c,c_\mathit{out}\}$}
    \end{array}
    $}[\graphlinewidth{0}\graphnodecolour{.7}\graphlinecolour{.7}]
\end{graph}

\end{center}
\caption{
    Basic membrane system $\BMS_0$.
    }
    \label{FIG-2}
\end{figure}

\paragraph{Basic membrane systems.}
Let $V$ be a finite alphabet of names of \emph{objects}
(or molecules) and let $\mu$ be a membrane structure of degree $m$.
A \emph{basic membrane system} (over $V$ and $\mu$) is a tuple
\[
    \BMS=(V,\mu,w_1^0,\ldots,w_m^0,R_1,\ldots, R_m)
\]
such that, for every membrane $i$,
$w_i^0$ is a multiset of objects from $V$, and
$R_i$ is a finite set of \emph{evolution rules}
associated with membrane (compartment) $i$.
Each evolution rule $r \in R_i$ is
of the form
$r:\lhs^r \to \rhs^r$,
where $\lhs^r$ (the left hand side of $r$) is a nonempty multiset over $V$,
and $\rhs^r$ (the right hand side of  $r$) is a
multiset over
\[
    V\cup \{a_\mathit{out}\mid a\in V\}
    \cup\{a_{{in}_j}\mid  a\in V ~and~(i,j)\in \mu\}\;.
\]
Here a symbol $a_{{in}_j}$ represents an object $a$
that is sent to a child node (compartment) $j$
and $a_\mathit{out}$ means that $a$ is sent to the parent node.
If $i$ is the root of
$\mu$ then no indexed object of the form $a_\mathit{out}$ belongs to $\rhs^r$.
A \emph{configuration} of $\BMS$ is
a tuple
\[
    C=(w_1,\ldots,w_m)
\]
of multisets of objects, and
$C_0=(w_1^0,\ldots,w_m^0)$ is the \emph{initial} configuration.
Figure~\ref{FIG-2} shows a basic membrane system
over the membrane structure depicted in Figure~\ref{FIG-1}.

A membrane system evolves from configuration to configuration
as a consequence of the application
of evolution rules.
There
are different \emph{execution modes} ranging from fully synchronous ---
as many applications of rules as possible --- to sequential
--- a single application of a rule at a time.
Here, similarly as in the case of \textsc{ptl}-nets,
we distinguish three modes,
all based on the notion of a vector multi-rule.

A \emph{vector multi-rule} of $\BMS$ is a tuple
$\mathbf{r}= \langle\mathbf{r}_1,\ldots, \mathbf{r}_m\rangle$
where, for each membrane $i$ of $\mu$, $\mathbf{r}_i$ is a multiset of rules from $R_i$.
For such a vector multi-rule, we denote by $\lhs_i^\mathbf{r}$
the multiset
\[
    \sum_{r\in R_i}\mathbf{r}_i(r)\cdot \lhs^r
\]
in which all
objects in the left hand sides of the rules in $\mathbf{r}_i$ are accumulated, and
by $rhs_i^\mathbf{r}$ the multiset
\[
    \sum_{r\in R_i}\mathbf{r}_i(r)\cdot \rhs^r
\]
of all
(indexed) objects in the right hand sides.
The first multiset specifies how many objects
are needed in each compartment
for the simultaneous execution of all the instances
of evolution rules in $\mathbf{r}$.

A vector multi-rule $\mathbf{r} $ of $\BMS$ is
\begin{itemize}
\item
    \emph{free-enabled}
    at a configuration $C$ if $\lhs_i^\mathbf{r}\leq w_i$, for each $i$.
\end{itemize}
Moreover, a free-enabled vector multi-rule
$\mathbf{r}= \langle\mathbf{r}_1,\ldots, \mathbf{r}_m\rangle$ is:
\begin{itemize}

\item
    \emph{max-enabled} if no $\mathbf{r}_i$ can be extended
    to a vector multi-rule which is free-enabled at $C$;  and

\item
    \emph{lmax-enabled} if no nonempty $\mathbf{r}_i$ can be extended
    to a vector multi-rule which is free-enabled at $C$.
\end{itemize}
For example, in Figure~\ref{FIG-2},
\begin{itemize}
\item
    $\langle\es,\es,\{\mathtt{r}_{31}\}  \rangle$ is not free-enabled;
\item
    $\langle \{\mathtt{r}_{11},\mathtt{r}_{12}\},\es,\es  \rangle$ is lmax-enabled but not max-enabled; and
\item
    $\langle \{\mathtt{r}_{11},\mathtt{r}_{12}\},\{\mathtt{r}_{21},\mathtt{r}_{22}\},\es  \rangle$ is max-enabled.
\end{itemize}

If $\mathbf{r}$ is free-enabled ($\mathit{free}$) at a configuration $C$,
then $C$ has in each membrane $i$ enough copies of objects
for the application of the multiset of evolution rules $\mathbf{r}_i$.
Maximal concurrency ($\mathit{max}$) requires that adding any extra rule makes $\mathbf{r}$
demand more objects than $C$ can provide.
Locally maximal concurrency ($\mathit{lmax}$) is similar but in this case only
those compartments which have rules in $\mathbf{r}$ cannot enable any more rules;
in other words, each compartment either uses no rule, or
uses a maximal multiset of rules.

The effect of  the rules is independent of the mode of execution
$\Mode\in\{\mathit{free}, \mathit{max},\mathit{lmax}\}$.
A  vector multi-rule $\mathbf{r} $ which is $\Mode$-enabled at
$C$ can \emph{$\Mode$-evolve} to a configuration $C'=(w'_1,\ldots w'_m)$
such that,
for each $i$ and object $a$:
\[
    w'_i(a) = w_i(a) -
    \lhs_i^\mathbf{r}(a) +
    \rhs_i^\mathbf{r}(a) +
    \rhs_{\mathit{parent}(i)}^\mathbf{r}(a_{{in}_i})
    +\!\!\!\!\!\!
    \sum_{i=\mathit{parent}(j)}\!\!\!\!\!\! \rhs_j^\mathbf{r}(a_\mathit{out})
\]
where $\rhs_{\mathit{parent}(i)}^\mathbf{r}=\es$ if $i$ is the root of $\mu$.
We denote this by $C\goesONE{\mathbf{r} }_\Mode C'$.
Moreover, an \emph{$\Mode$-computation} is a finite sequence of $\Mode$-evolutions
starting from the initial configuration; any configuration which can be obtained
through such a computation is called \emph{$\Mode$-reachable}.
For the basic membrane system depicted in Figure~\ref{FIG-2} we have, for example:
\[
    C_0
    \xrightarrow{\langle \{\mathtt{r}_{11},\mathtt{r}_{12}\},\es,\es  \rangle}_\mathit{lmax}
    (\{a,b\},\{a,b,c,c,c\},\{a\})
    \xrightarrow{\langle \es, \{\mathtt{r}_{21},\mathtt{r}_{22}\},\es  \rangle}_\mathit{lmax}
    (\{a,b\},\{a,b,c,c\},\{a\})
    \;.
\]

Let $\Mode\in\{\mathit{free}, \mathit{max},\mathit{lmax}\}$ be a mode
of execution of a basic membrane system $\BMS$.
Then
the \emph{$\Mode$-concurrent reachability graph of $\BMS$} is given by:
\[
    \CRG_\Mode(\BMS)
    =
    \Big(~\STEP{C_0}_\Mode~,~
   \big\{(C,\mathbf{r}_1+\ldots+\mathbf{r}_m,C')
    \mid C\in \STEP{C_0}_\Mode\;\wedge\;
    C\xrightarrow{\langle\mathbf{r}_1,\ldots, \mathbf{r}_m\rangle}_\Mode C'
    \big\}~,~C_0~\Big) \;,
\]
where $\STEP{C_0}_\Mode$ is the set of all
$\Mode$-reachable configurations
which are the nodes of the graph; $C_0$ is the initial node;
and the arcs between the nodes are labelled by
multisets of
evolution rules
involved in the $\Mode$-executed vector multi-rules.\footnote{
Though it may be that rules from different membranes are the same in terms
of the multisets defining their left hand and right hand sides,
we assume here that evolution rules associated with different membranes
can be distinguished, \eg by giving them each their own name (an injective
label).
}
Similarly as in the case of \textsc{ptl}-nets,
concurrent reachability graphs capture completely
the dynamic behaviour of basic membrane system
evolving according to the chosen mode of execution.

\section {Membrane Systems and Petri Nets}

There is a natural way of translating a basic
membrane system $\BMS=(V,\mu,w_1^0,\ldots,w_m^0,R_1,\ldots, R_m)$
over a membrane structure $\mu$
into a behaviourally equivalent \textsc{ptl}-net $\PTL (\BMS)=(P,T,W,\ell,M_0)$
spanned
over the same membrane structure.
In the constructed net, places represent objects present inside compartments,
and transitions represent evolution rules.
Both places and transitions are associated with membranes
and this information is represented by the location mapping.

The constructed \textsc{ptl}-net
$\PTL (\BMS)$ has a separate
place $\mathtt{p}_j^a$ with $\ell(\mathtt{p}_j^a)=j$,
for each object $a$ and membrane $j$, and
a separate transition $\mathtt{t}_i^r$ with $\ell(\mathtt{t}_i^r)=i$,
for each rule $r$ in compartment $i$.

The initial marking inserts
$w_j^0(a)$
tokens into each place $\mathtt{p}_j^a$.
The connectivity between
transition $t=\mathtt{t}_i^r$ and place $p=\mathtt{p}_j^a$ is given by:
\[
    W(p,t)=
    \makebox[140pt][l]
    {$
    \left\{
    \begin{array}{ll}
    \lhs^r(a) & \mbox{if $i=j$}
    \\
    0 & \mbox{otherwise}\;,
    \end{array}
    \right.
    $}
\]
as well as:
\[
    W(t,p)=
    \makebox[140pt][l]
    {$
    \left\{
    \begin{array}{ll}
    \rhs^r(a) & \mbox{if $i=j$}
    \\
    rhs^r(a_\mathit{out}) & \mbox{if $j=\mathit{parent}(i)$}
    \\
    \rhs^r(a_{{in}_j}) & \mbox{if $i=\mathit{parent}(j)$}
    \\
    0 & \mbox{otherwise}\;.
    \end{array}
    \right.
    $}
\]
Figure~\ref{FIG-3aa} shows the result of the above translation for the
basic membrane system in Figure~\ref{FIG-2}.
Note that it immediately follows from the construction that
the \textsc{ptl}-net $\PTL (\BMS)$ is spanned over $\mu$.

The \textsc{ptl}-net $\PTL (\BMS)$ provides a
\emph{faithful} representation of the behaviour
of the   basic membrane system $\BMS$.
To capture this very close relationship,
we define two bijective mappings, $\nu$ and $\rho$,
which allow us to move between $\BMS$ and $\PTL (\BMS)$:
\begin{itemize}
\item
    for every marking $M$
    of $\PTL (\BMS)$,
    $\nu(M)=(w_1,\ldots,w_m)$
    is the
    configuration of $\BMS$
    given by $w_i(a)=M(\mathtt{p}_i^a)$,
    for every object $a$ and every $i$.
\item
    for every step $U$
    of $\PTL (\BMS)$,
    $\rho(U)=\langle\mathbf{r}_1,\ldots, \mathbf{r}_m\rangle$
    is the
    vector multi-rule of $\BMS$
    given by
    $\mathbf{r}_i(r)=U(\mathtt{t}_i^r)$,
    for every rule $r\in R_i$ and every $i$.
\end{itemize}
It is then possible to establish a direct relationship
between (the operation of) the original
membrane system and the \textsc{ptl}-net
resulting from the above translation at the system level:
\begin{equation}
\label{result-main}
\begin{array}{lcl}
    C\goesONE{\mathbf{r}}_\Mode C'
    &
    \Longrightarrow
    &
    \nu^{-1}(C)\;\STEP{\rho^{-1}(\mathbf{r})}_\Mode \;\nu^{-1}(C')
\\[2mm]
    M\STEP{U}_\Mode M'
    &
    \Longrightarrow
    &
    \nu(M)\;\goesONE{\rho(U)}_\Mode \;\nu(M')
\end{array}
\end{equation}
for all modes of execution
$\Mode\in\{\mathit{free}, \mathit{max},\mathit{lmax}\}$,
configurations $C$ of $\BMS$ and markings $M$ of $\PTL (\BMS)$.
Together with $\nu(M_0)=C_0$,
this result means that the
$\Mode$-step sequences of $\PTL (\BMS)$
faithfully represent
$\Mode$-computations of $\BMS$, and the same applies to markings and
configurations.
Crucially, we obtain

\begin{theorem}
\label{th-crg}
    For each $\Mode\in\{\mathit{free},\mathit{max},\mathit{lmax}\}$,
\[
    \CRG_\Mode(\PTL (\BMS)) \sim_{\phi,\nu}\CRG_\Mode(\BMS) \;,
\]
    where the mapping $\nu$ is defined as above, and
    $\phi(\mathtt{t}_i^r)=r$, for every transition $\mathtt{t}_i^r$
    of $\PTL (\BMS)$.
\end{theorem}

The above theorem captures the very tight behavioural correspondence
between $\BMS$ and $\PTL (\BMS)$,
allowing to apply analytical
techniques developed for Petri nets in the analysis of membrane systems.
For example, one can employ
the invariant analysis based on linear algebra~\cite{inv},
or use the
causality semantics approach of Petri nets based on occurrence nets,
as first outlined in~\cite{KKR06}.
In this paper, we show how techniques used
to synthesise Petri nets could be
employed in order to construct basic membrane systems from their
intended behaviours as represented by step transition systems.
First, however, we provide a translation from
\textsc{ptl}-nets spanned over membrane structures to basic membrane
systems.

Let $\PTL =(P,T,W,\ell,M_0)$ be a \textsc{ptl}-net
spanned over a membrane structure $\mu$.
For such a \textsc{ptl}-net, we  construct the corresponding basic
membrane system $\BMS(\PTL)$ over $\mu$ in the following way:
\begin{itemize}
\item
    $P$ is the set of objects;

\item
    the initial configuration is $\nu'(M_0)$ where, for every marking $M$ of $\PTL$,
\[
    \nu'(M)=
     (M|_{\ell^{-1}(1)\cap P},\ldots,M|_{\ell^{-1}(m)\cap P})\;;
\]

\item
    each transition $t\in T$ with $\post{t}=\{p^1,\ldots,p^k\}$ has
    a corresponding evolution rule $\phi'(t)$ of the form
    $t:\pre{t}\to\{a_1,\ldots,a_k\}$
    where, for $i=1,\ldots,k$,
\[
    a_i
    =\left\{
    \begin{array}{ll}
    p^i
    &\mbox{if } \ell(p^i)=\ell(t)
    \\[1mm]
    p^i_\mathit{out}
    &\mbox{if } \ell(p^i)=\mathit{parent}(\ell(t))
    \\[1mm]
    p^i_{\mathit{in}_{\ell(p^i)}}
    &\mbox{if } \ell(t)=\mathit{parent}
    (\ell(p^i))\;
    \end{array}
    \right.
\]

\item
    for each membrane $i\in\mu$, the set of evolution
    rules is given by $R_i=\{\phi'(t)\mid \ell(t)=i\}$.
\end{itemize}

Again, the translation results in a very close
behavioural correspondence.

\begin{theorem}
\label{th-crg1}
    For each $\Mode\in\{\mathit{free},\mathit{max},\mathit{lmax}\}$,
\[
    \CRG_\Mode(\PTL)\sim_{\phi',\nu'}\CRG_\Mode(\BMS(\PTL))\;,
\]
    where the mappings
    $\phi'$ and $\nu'$ are defined as above.
\end{theorem}

It follows from Theorems~\ref{th-crg} and \ref{th-crg1}
that the problem of synthesis of basic membrane systems from
step transition systems is equivalent to the problem of synthesis of
\textsc{ptl}-nets spanned over membrane structures.
It therefore suffices to solve the latter, and in the next section
we describe a solution based on the notion of a
region of a step transition system.

\section{Synthesising nets corresponding to membrane systems}
\label{sectdd-2-3}

The Petri net synthesis problem we consider is formulated as follows.

\begin{problem}
\label{problem-2}
    Given are a finite set $T$, a membrane structure $\mu$, a mapping
    $\ell:T\to\mu$,
    $\Mode\in\{\mathit{free},\mathit{max},\mathit{lmax}\}$,
    and $\TS=(Q,\mathcal{A},q_0) $ which is a finite step transition system over $T$.
\\
    Construct a \textsc{ptl}-net $\PTL=(P,T,\ell,M_0)$ spanned
    over   $\mu$
    such that $\CRG_\Mode(\PTL)\sim\TS$, and $\ell$ is an extension of
    the mapping defined for $T$.
\end{problem}

As demonstrated in~\cite{DKPKY08}, synthesis problems
like Problem~\ref{problem-2} can be solved
using techniques coming from the theory of
regions of transition systems (see, e.g., \cite{BD97,ER90,M92}).
Intuitively, a region represents a
single place in a hypothetical net generating the given transition
system.
Regions are used both to check whether a net
satisfying the conditions can be constructed and, if the answer
turns out to be positive,
to construct such net.

In this particular case, a \emph{region}
of the step transition system
$\TS$ consists of three mappings
\begin{equation}
\label{eq-region}
    \mathit{reg}
    =
    \big(~\sigma:Q\rightarrow \nsymbol~,~
    \imath:T\rightarrow \nsymbol~,~
    \omega:T\rightarrow \nsymbol~\big)
\end{equation}
such that, for every arc
$q\xrightarrow{\alpha}q'$ of $\TS$,
\begin{equation}
\label{reg-ooo}
    \sigma(q) \geq \omega(\alpha)
    ~~~\mbox{and}~~~
    \sigma(q')
    =\sigma(q)-\omega(\alpha)+\imath(\alpha)\;.
\end{equation}
Here
$\omega(\alpha) = \sum_{t\in T}\alpha (t) \cdot \omega(t)$
and similarly
$\imath(\alpha) = \sum_{t\in T}\alpha (t) \cdot \imath(t)$.
In a region of the form (\ref{eq-region}) representing a place $p$,
$\sigma(q)$ is
the number of tokens in $p$ in the marking corresponding to the
node $q$, $\omega(t)$ represents the weight of the arc from $p$
to transition $t$, and $\imath(t)$ represents the weight of the arc from
$t$ to~$p$.
It is then natural to require in (\ref{reg-ooo})
that $p$ contains enough tokens not to block a step $\alpha$
executed at $q$, and also to ensure that the number of tokens in
$p$ before and after executing $\alpha$ is consistent
with the total arc weight of the step $\alpha$ in relation to $p$.

In the case of Problem~\ref{problem-2},
one also needs to take into account the
fact that the target \textsc{ptl}-net must be spanned over $\mu$.
This imposes additional constraints on allowed regions (places)
and the location mapping $\ell$.
We call a region $\mathit{reg}$
as in (\ref{eq-region})
with a location $\ell(\mathit{reg}) \in \mu$
\emph{compatible with}
the membrane structure $\mu$ if the following hold, for every $t\in T$:
\begin{itemize}
\item
    if $\omega(t)>0$
    then
    $\ell(t)=\ell(\mathit{reg})$; and

\item
    if $\imath(t)>0$
    then
    $\ell(t)=\ell(\mathit{reg})$ or
    $(\ell(t),\ell(\mathit{reg}))\in\mu$ or
    $(\ell(\mathit{reg}),\ell(t))\in\mu$.
\end{itemize}
The set of all such regions will be denoted by $\mathbf{P}_\mu$.
Note that if $\mathit{reg}$ is such that $\omega(t)>0$, for at least one $t\in T$,
then $\ell(\mathit{reg})$ is uniquely determined; otherwise we always choose
$\ell(\mathit{reg})$ to be the membrane which is higher up in the tree
structure of $\mu$ than any other suitable candidate. As a result,
we can leave $\ell(\mathit{reg})$ implicit.

Finally, Problem~\ref{problem-2} should be feasible in the sense that
the transition system $\TS$ can be realised by a suitable net. There
are two necessary and sufficient conditions for realisability
(see~\cite{DKPKY08,MKMPK09}):
\begin{itemize}
\item
    \emph{state separation}:
    for every pair of distinct states of the transition system there is
    a region (a marked place) distinguishing between them;
    and
\item
    \emph{forward closure}:
    there are sufficiently many places defined by regions of the
    transition system to disallow steps not present in the transition system.
\end{itemize}
First we describe how all places can be found that potentially
provide a solution to Problem~\ref{problem-2};
in other words, all the regions (\ref{eq-region})
of the transition system $\TS$ which are compatible with $\mu$.

\paragraph{Finding compatible regions.}

Let $T$, $\mu$, $\ell:T\to\mu$ and $\TS=(Q,\mathcal{A},q_0)$
be as in Problem~\ref{problem-2}.
Assume that
$Q=\{q_0,\ldots,q_h\}$ and $T=\{t_1,\ldots,t_n\}$.
We use three vectors of non-negative variables:
\begin{align*}
    \xvec&=x_0\ldots x_h&
    \yvec&=y_1\ldots y_n&
    \zvec&=z_1\ldots z_n\;.
\end{align*}
We also denote $\pvec=\xvec\yvec\zvec$ and
define a homogeneous linear system
\[
\mathcal{P}~:~
\left\{
\begin{array}{ll}
    x_i\geq \alpha\cdot\zvec \\
    x_j=x_i + \alpha\cdot(\yvec-\zvec)~~~~~~~~
\end{array}
    \mbox{for all $q_i\xrightarrow{\alpha}q_j$ in $\TS$}
\right.
\]
where $\alpha\cdot\zvec$ denotes
$\alpha(t_1)\cdot z_1+\cdots+\alpha(t_n)\cdot z_n$
and similarly for $\alpha\cdot(\yvec-\zvec)$.

The regions (\ref{eq-region}) of
$\TS$ are then determined by the
integer solutions $\pvec$ of the system $\mathcal{P}$ assuming that,
for $0\leq i\leq h$
and $1\leq j\leq n$,
\begin{align*}
    \sigma(q_i)&=x_i
    &
    \imath(t_j)&=y_j
    &
    \omega(t_j)&=z_j
\end{align*}

The set of rational solutions of $\mathcal{P}$
forms a polyhedral cone in
$\mathbb{Q}^{h+2n+1}$.
As described in~\cite{Che65}, one can effectively compute
\underline{finitely} many integer generating rays
$\pvec^1,\ldots,\pvec^k$
of this cone such that any
integer
solution $\pvec$ of $\mathcal{P}$
can be expressed as a linear combination of the rays
with non-negative rational coefficients:
\[
    \pvec=\sum_{l=1}^k c_l \cdot \pvec^l\;.
\]
Such rays $\pvec^l$ are fixed
and (some of them) turned into net places if
Problem~\ref{problem-2}
has a solution.
More precisely, if $\pvec^l$ is included in the constructed net, then
\begin{align}
\label{eq-connect}
    M_0(\pvec^l)
    &= x^l_0
    &
    W(\pvec^l,t_i)
    &= z^l_i
    &
    W(t_i,\pvec^l)
    &= y^l_i\;,
\end{align}
where $M_0$ is the initial marking of the target net, and $t_i\in T$.

Clearly, not all such rays can be considered for the
inclusion in the net being constructed, as the corresponding
regions have to be compatible with $\mu$.
We therefore ensure through a simple check
that the generating rays $\pvec^1,\ldots,\pvec^k$
are compatible with $\mu$,
deleting in the process those which are not.
Note that any $\pvec\in\mathbf{P}_\mu$ is a non-negative linear
combination of rays compatible with $\mu$.

Having found the generating rays compatible with $\mu$,
we proceed to check whether Problem~\ref{problem-2}
has any solutions at all.

\paragraph{Checking state separation.}
Let $\TS=(Q,\mathcal{A},q_0)$ be as in Problem~\ref{problem-2}.
We take in turn each pair of distinct
states, $q_i$ and $q_j$, of $Q$ and
decide whether there exists
$\pvec=(\sigma,\imath,\omega)\in\mathbf{P}_\mu$ with
coefficients $c_1,\ldots,c_k$ such that $\sigma(q_i)=x_i\neq x_j=\sigma(q_j)$.
Since the latter is equivalent to
\[
    \sum_{l=1}^k c_l \cdot x_i^l \neq \sum_{l=1}^k c_l \cdot x_j^l\;,
\]
one can simply check whether there
exists at least one $\pvec^l$ (called a \emph{witness}~\cite{DR-96})
such that $x_i^l\neq x_j^l$.

\paragraph{Checking forward closure.}
Again, let $\TS=(Q,\mathcal{A},q_0)$ be as in Problem~\ref{problem-2},
and $\Mode\in\{\mathit{free},\mathit{max},\mathit{lmax}\}$.
First, we take in turn
each state $q_i$ of $Q$, and
calculate the set of \emph{region enabled} steps,
denoted by $\RS_{q_i}$.
Intuitively, region enabled steps are those that cannot be disabled (or blocked) by
compatible regions.

To build $\RS_{q_i}$ one
only needs to consider nonempty steps $\alpha$ with
$|\alpha|\leq m\cdot\mathit{Max}$,
where $\mathit{Max}$ is the maximum size
of steps labelling arcs in $\TS$, and $m$ is the number of membranes of $\mu$.
The reason is that, for each membrane $i\in\mu$
there exists a compatible region
$(\sigma,\imath,\omega)\in\mathbf{P}_\mu$ (called a \emph{witness})
such that $\sigma(Q)=\{\mathit{Max}\}$ and, for every $t\in T$,
\[
    \omega(t)=\iota(t)=
    \left\{
    \begin{array}{ll}
    1~~&\mbox{if }\ell(t)=i
    \\
    0~~&\mbox{otherwise}\;.
    \end{array}
    \right.
\]
Taken together, all such regions block any step  $\alpha$ with
$|\alpha|> m\cdot\mathit{Max}$.

For each nonempty step $\alpha$ with $|\alpha|\leq m\cdot\mathit{Max}$ it is the case that
$\alpha\notin\RS_{q_i}$ \emph{iff} for some $\pvec\in\mathbf{P}_\mu$
with coefficients $c_1,\ldots,c_k$ we have
$x_i<\alpha\cdot\zvec$. Since the latter is equivalent to
\[
    \sum_{l=1}^k c_l\cdot(x_i^l- \alpha\cdot \zvec^l) <0\;,
\]
one simply checks
whether there exists at least one $\pvec^l$ (again called a \emph{witness}) such
that $x_i^l-\alpha\cdot \zvec^l<0$.

Having determined the region enabled steps, in order to establish forward closure
we need to verify that, for every state $q\in Q$,
\[
    \AS_q=
    \left\{
    \begin{array}{ll}
    \RS_q
    &\mbox{if }\Mode=\mathit{free}
    \\[1mm]
    \{\alpha\in \RS_q\mid
      \neg \exists t\in T: ~\alpha+\{t\}\in \RS_q \}
    &\mbox{if }\Mode=\mathit{max}
    \\[1mm]
    \{\alpha\in \RS_q\mid
      \neg \exists t\in T: ~\alpha+\{t\}\in \RS_q~
      \wedge~
    \ell(t)\in\ell(\alpha)\}
    &\mbox{if }\Mode=\mathit{lmax}\;.
    \end{array}
    \right.
\]

\paragraph{Constructing the solution net.}

If the  above checks for the
feasibility of Problem~\ref{problem-2} are successful, one can
construct a solution \textsc{ptl}-net spanned over $\mu$ by taking all the
witness rays and regions,
and treating them as places in the way indicated in (\ref{eq-connect}).
The resulting net $\PTL$  satisfies
\[
    \CRG_\Mode(\PTL)\sim\TS\;.
\]

\section{Concluding remarks}

We have described how one can adapt a solution
to the Petri net synthesis problem based on regions of step
transition systems, so that the resulting
method can be used to construct basic membrane systems
with a specific behaviour.
Moreover, there are other synthesis results developed for
Petri nets which can be employed to extend the proposed solution
in several directions, two of which are briefly mentioned below.

In Problem~\ref{problem-2} it is assumed that the
association of transitions with
membranes is given. This can be relaxed and one can
aim at synthesising membrane systems without such an association, or even
without being given a membrane structure (in such a case, the synthesis
procedure should construct a membrane structure as well).
For such a modification, there already exist  results which
can be used to develop a solution. More precisely,
the method of `discovering' localities in~\cite{MKMPK09} works for \textsc{ptl}-nets
with localised conflicts (where
transitions which share an input place are co-located).
Since all \textsc{ptl}-nets spanned over membrane structures have localised conflicts,
the result in~\cite{MKMPK09} can be adapted to work for basic membrane systems.

Evolution rules of membrane systems are often equipped with promoters and inhibitors.
Both features have direct counterparts in Petri nets in the form of activator and
inhibitor arcs, and
suitable translations between membrane systems and Petri nets can be developed
as described in~\cite{our-handbook,KKR11}.
Moreover, the synthesis technique based on regions
of step transition systems works also for \textsc{ptl}-nets extended with
activator and inhibitor arcs~\cite{MKMPK08}.
In fact, there is a general  setting of so-called
$\tau$-nets and corresponding $\tau$-regions~\cite{BD97,DKPKY08}.
Here the parameter $\tau$ is a general and convenient way of capturing
different types of connections (arcs and their combinations)
between places and transitions, removing the need to
re-state and re-prove the key results
every time a new kind arcs is introduced.
Note that the recently introduced
\textsc{set}-nets~\cite{KKR11a,KKR11}
(with qualitative rather than quantitative resource management)
and set membrane systems~\cite{KK-SET-MS} can be
treated within the general theory of $\tau$-net synthesis
based on regions of transition systems~\cite{KKPKR}.

\subsection*{Acknowledgements}
This paper is based on an invited talk presented at the
6th Workshop on Membrane Computing and
Biologically Inspired Process Calculi (\textsc{MeCBIC}),
8th September 2012, Newcastle upon Tyne, United Kingdom.
The reported research was supported by
the \textsc{Epsrc} \textsc{Gaels} project.

\end{document}